\documentclass{article}

\usepackage[english]{babel}

\usepackage[letterpaper,top=2cm,bottom=2cm,left=3cm,right=3cm,marginparwidth=1.75cm]{geometry}

\usepackage{amsmath}
\usepackage{graphicx}
\usepackage{amssymb}
\usepackage[colorlinks=true, allcolors=blue]{hyperref}
\RequirePackage{subcaption} 

\title{An Information Bottleneck Asset Pricing Model}
 \author{Che Sun \thanks {PBC School of Finance, Tsinghua University, sunch@pbcsf.tsinghua.edu.cn}} 
\begin{document}
\maketitle

\begin{abstract}
Deep neural networks (DNNs) have garnered significant attention in financial asset pricing, due to their strong capacity for modeling complex nonlinear relationships within financial data. However, sophisticated models are prone to over-fitting to the noise information in financial data, resulting in inferior performance. To address this issue, we propose an information bottleneck  asset pricing model that compresses data with low signal-to-noise ratios to eliminate redundant  information and retain the critical information for asset pricing. Our model imposes constraints of mutual information during the nonlinear mapping process. Specifically, we progressively reduce the mutual information between the input data and the compressed representation while increasing the mutual information between the compressed representation and the output prediction. The design ensures that irrelevant information, which is essentially the noise in the data, is forgotten during the modeling of financial nonlinear relationships without affecting the final asset pricing. By leveraging the constraints of the Information bottleneck, our model not only harnesses the nonlinear modeling capabilities of deep networks to capture the intricate relationships within financial data but also ensures that noise information is filtered out during the information compression process.
\end{abstract}

\section{Introduction}

The field of asset pricing has long been concerned with understanding the determinants of asset values and the factors that influence the returns on those assets\cite{cochrane2009asset,campbell2000asset,brunnermeier2021perspectives}. Central to this endeavor is the concept of the pricing kernel, also known as the stochastic discount factor (SDF)\cite{hansen1997assessing}, which is a theoretical construct that represents the marginal rate of substitution between consumption in different periods. The pricing kernel encapsulates all the information about the future payoffs of assets and the uncertainty associated with those payoffs, as perceived by market participants.

Traditional asset pricing models, such as the Capital Asset Pricing Model (CAPM) \cite{sharpe1964capital} and the Fama-French three-factor model \cite{fama1993common}, rely on a set of predefined risk factors to explain the cross-sectional variation in asset returns. These models are parametric in nature, meaning they make specific assumptions about the form of the pricing kernel and the sources of risk. However, the complexity and uncertainty of financial markets often defy such simplifications, leading to models that may not fully capture the underlying dynamics of asset prices.

The recent strides in the field of machine learning (ML) have introduced a novel, data-driven paradigm to the realm of dynamic factor models\cite{kozak2020shrinking,gu2021autoencoder,gu2020empirical}. Owing to their enhanced ability to discern intricate patterns within market data, certain ML methodologies are capable of autonomously extracting latent factors from this data, potentially surpassing the efficacy and applicability of conventional approaches in real-world market contexts. However, prevailing ML solutions are not immune to a critical challenge, namely, the low signal-to-noise ratio characteristic of stock data. The proliferation of noisy data can significantly impede the learning processes of ML-based models, thereby compromising the efficacy of the latent factors they extract. This issue indeed poses a formidable impediment to the development of an effective factor model aimed at asset pricing.

To break this barrier, we introduce an Information Bottleneck-based Asset Pricing Model, which is designed to distill data characterized by low signal-to-noise ratios. This model is adept at filtering out redundant information while preserving the pivotal elements necessary for asset valuation. Our approach incorporates mutual information constraints within the nonlinear mapping framework. Specifically, the model is engineered to decrement the mutual information between the input dataset and its compressed counterpart, concurrently enhancing the mutual information between the compressed representation and the ensuing asset pricing predictions. This architecture ensures that extraneous information, which is fundamentally noise within the dataset, is effectively discarded during the modeling of financial nonlinear dynamics, without compromising the integrity of the final asset pricing outcomes. By exploiting the Information Bottleneck constraints, our model not only capitalizes on the nonlinear modeling prowess of deep neural networks to capture the nuanced interdependencies within financial datasets but also ensures that noise is effectively mitigated during the information compression phase.
\section{Related Work}

\subsection{Factor Model}

Factor models can be taxonomized into two principal categories: static models and dynamic models. Within the domain of static factor models, the exposure of stocks to underlying factors is posited to be invariant over time. The progenitor of static factor models is the Capital Asset Pricing Model (CAPM)\cite{sharpe1964capital}, which introduces the market factor and posits that discrepancies in stock returns can be ascribed to varying sensitivities to this market factor. Subsequently, in a groundbreaking study,  Fama and  French observed that firm size and value are significant determinants in explaining the divergence in expected stock returns. Building upon the foundational tenets of CAPM, they augmented the model by incorporating size and value risk factors, thereby formulating the renowned Fama -French three-factor model \cite{fama1993common}. This seminal contribution expanded the explanatory power of asset pricing models by acknowledging the role of additional risk factors beyond the market premium. Recent research \cite{fama2020comparing} indicates that dynamic models with time-varying factor loadings  achieve better asset pricing performance than static methods, so dynamic models become increasingly popular.

\subsection{Asset Pricing with Machine Learning}
In their seminal working paper, Gu et al. \cite{gu2020empirical} conducted a comparative analysis of the deployment of machine learning methodologies in the context of empirical asset pricing challenges, with a particular focus on the quantification of asset risk premiums. Their findings revealed that investors employing machine learning-based predictions can garner substantial economic advantages, with certain methodologies, such as tree-based \cite{shorish2005shaking} and neural network models\cite{jiang2023re,duan2022factorvae}, outperforming established regression strategies by a factor of two. The superior performance was attributed to the capacity of these models to facilitate interactions among non-linear predictive factors, a capability that is absent in more traditional approaches\cite{chen2024deep}.

Machine learning techniques have been instrumental in the analysis of conditional factor models, where they are leveraged to discern stochastic discount factors and to systematically assess and appraise prevailing asset pricing frameworks. Illustratively, Fama and French \cite{fama1993common} introduced an asset pricing model predicated on linear factor dynamics, whereas Giglio et al. \cite{giglio2021thousands} advanced a novel methodology for examining the cross-sectional returns within a linear paradigm, taking into account an extensive array of conditional information.

Furthermore, machine learning has been instrumental in addressing prediction challenges within empirical asset pricing. Freyberger et al. \cite{freyberger2020dissecting} employed non-parametric techniques to dissect the underlying features, while Gu et al. \cite{gu2021autoencoder} proposed an autoencoder asset pricing model that extends the scope of linear conditional factor models to encompass non-linear factor dynamics. The refinement of risk premium measurement through machine learning has streamlined the exploration of asset pricing economic mechanisms and underscored the pivotal role of machine learning in financial innovation. This is evident in the work of Gu et al. \cite{gu2021autoencoder}, who underscore the transformative impact of machine learning in finance. In recent years, the proliferation of data availability and advancements in computational capabilities have augmented the prominence of machine learning within the financial literature. Harvey and Liu  \cite{harvey2021lucky} deliberated on the prevalence of spurious (and omitted) discoveries in financial economics, while Giglio and Xiu \cite{giglio2021asset} investigated the implications of omitted factors on asset pricing models.

\subsection{Information Bottleneck}

The information bottleneck method has been proposed by  Tishby et al. \cite{tishby2000information}, aiming to balance the relationship between data compression and information retention by optimizing information retention during the data compression process\cite{saxe2019information,wu2020graph,chechik2003information}. The application of information bottleneck theory in the field of deep learning is becoming increasingly widespread, especially in explaining the representation mechanism of deep neural networks. Tishby et al. \cite{tishby2015deep} further explored the relationship between deep learning and information bottleneck theory, proposing that deep learning can be seen as an information bottleneck process aimed at compressing noisy data while preserving the information represented by the data. The information bottleneck method is used to improve the generalization ability of the model. By controlling the amount of information in the model parameters, i.e. the model complexity, overfitting and underfitting problems can be avoided. In addition, the combination of information bottleneck method and PAC Bayes boundary theory can optimize the upper bound of test loss\cite{wang2021pac}. As research deepens, various variants of information bottleneck methods have emerged, such as Deterministic Information Bottleneck (DIB) \cite{strouse2017deterministic} and Elastic Information Bottleneck (EIB)\cite{ni2022elastic}. EIB balances the trade-off between source generalization gap and representation difference by interpolating between IB and DIB to achieve better domain adaptation results. The information bottleneck method has also been applied in multi-task learning\cite{yan2023multitask} and transfer learning \cite{hu2024survey} to improve the adaptability and generalization ability of models across different tasks and domains.

\section{Method}

We use the Autoencoder asset pricing model \cite{gu2021autoencoder} as the baseline factor model and incorporate regularization constraints of the information bottleneck. Below, we will introduce them separately.

\subsection{Autoencoder Asset Pricing Model}
The autoencoder constitutes a sophisticated mechanism for dimensionality reduction, designed to extract a lower-dimensional representation of the input data by traversing it through a central, hidden layer that possesses a significantly reduced number of neurons relative to the dimensionality of the cross-sectional features. This process aligns with the underlying ethos of Principal Component Analysis (PCA), yet the autoencoder transcends the limitations of PCA by accommodating non-linear transformations in the data compression paradigm. In the ensuing section, we shall delve into an exploration of the interplay between autoencoders and PCA, examining their respective roles and the synergies that may be harnessed in the context of data dimensionality reduction.

In the autoencoder asset pricing model, the return models assume a linear latent factor with dynamic loading:
\begin{equation}
r_t=\beta_{t-1} f_t+u_t \label{eq1}
\end{equation}
where $r_t$ is the vector of returns exceeding the risk-free rate, $f_t$ is the vector of factor returns, $u_t$ is the vector of specificity error (uncorrelated), and $\beta_{t-1}$ is the matrix of factor loadings.

The autoencoder model designates the $\beta_{t-1}$ as the neural network model of the lagged company feature $z_{t-1}$, that is, the expression of the nonlinear function is
\begin{equation}
\beta_{t-1}=g^\theta\left(z_{t-1}\right)
\end{equation}
where $g^\theta$ is a  fully connected network with the parameter $\theta$. The mathematical expression of the factor $f_t$  is
\begin{equation}
f_t=b+W r_t
\end{equation}
where $W$ is the Learnable parameter matrix and $b$ is the Learnable parameter vector of the factor network. Finally, the ``dot operation'' multiplies the matrix output of the beta network $g^\theta(\cdot)$ with the output of the factor network to produce the final model fit for each asset return. All model parameters are optimized using the stochastic gradient descent algorithm with the backpropagation algorithm. The optimization objective of the autoencoder asset pricing model is 
\begin{equation}
\min _{\theta, b, W} \sum_{t=1}^T\left\|r_t-\beta_{t-1} f_t\right\|^2 \label{eq4}
\end{equation}
The autoencoder model incorporates a penalty term similar to LASSO to prevent model overfitting, given by

\begin{equation}
\min _{\theta, b, W} \sum_{t=1}^T\left\|r_t-\beta_{t-1} f_t\right\|^2 +\lambda\phi(\theta, b, W)
\end{equation}
where $\lambda$ is a trade-off hyperparameter, and 
\begin{equation}
\phi(\theta, b, W)= \sum_j\left|\theta_j\right|+\left|W\right|+\left|b\right|
\end{equation}
In this sense, LASSO applies sparsity to weight parameters, encouraging irrelevant weights to disappear.

\subsection{Information Bottleneck}
\subsubsection{mutual information }
In probability theory and information theory, the mutual information (MI) of two random variables is a measure of the mutual dependence between the two variables. More specifically, it quantifies the ``amount of information'' (in units such as Shannons, commonly called bits) obtained about one random variable through observing the other random variable. The concept of mutual information is intricately linked to that of entropy of a random variable, a fundamental notion in information theory that quantifies the expected ``amount of information'' held in a random variable. The expected amount of information is entropy
\begin{equation}
H(X)=-\sum_x p\left(x_i\right) \log p\left(x_i\right)
\end{equation}
where $X$ is a random variable. 
The formula for calculating conditional entropy is
\begin{equation}
\begin{aligned}
H(Y \mid X) & =\sum_{x \in X} p(x) H(Y \mid X=x) \\
& =-\sum_{x \in X} p(x) \sum_{y \in Y} p(y \mid x) \log p(y \mid x) \\
& =-\sum_{x \in X} \sum_{y \in Y} p(x, y) \log p(y \mid x)
\end{aligned}
\end{equation}
Therefore, the calculation of mutual information is
\begin{equation}
I(X ; Y)=\sum_{x, y} p(x, y) \log \frac{p(x, y)}{p(x) p(y)}=H(X)-H(X \mid Y)=H(Y)-H(Y \mid X)
\end{equation}

Information Bottleneck is an application or exception of Rate telling Theory. In deep networks, we aim to compress $X$ as much as possible into the representation $Z$, while meeting two requirements: first, to compress as much information as possible that is irrelevant to the learning objective $Y$ (such as the class label to be predicted) in the sample $X$, and second, to minimize the distortion of constructing the objective Y using Z. The degree of compression is represented by the  mutual information  of $X$ and $Z$, that is, $I (X; Z)$; while distortion is represented by the negative MI of $Y$ and $Z$, that is, $- I (Y; Z)$. Taking one of these two objectives as the optimization objective and the other as the regularization constraint, the Lagrange multiplier is used to construct a minimization optimization problem such as $\min I (X; Z) - \lambda I (Y; Z)$. Intuitively speaking, there exists a trade-off, and the second item is to find the most relevant information between $X$ and $Y$; And the first one is to make the relevant information as short as possible. So the final result is to find the information in $X $ that is most relevant to $Y$.

\subsection{Information Bottleneck Asset Pricing Model}
Different from the regularization based on simple sparsity assumptions, we propose a regularization method with information bottleneck constraints to eliminate redundant information and retain the critical information for asset pricing. 
Therefore, the objective function in Eq. (\ref{eq4}) incorporating information bottleneck constraints becomes 
\begin{equation}
\min _{\theta, b, W} \sum_{t=1}^T\left\|r_t-\beta_{t-1} f_t\right\|^2-\left(I\left(r_t ; \beta_{t-1}\right)-\lambda I\left(z_{t-1} ; \beta_{t-1}\right)\right)
\end{equation}

Considering that factors loadings and cross-sectional returns are high-dimensional, mutual information is difficult to calculate directly. Therefore, We perform variational approximation to solve the information bottleneck. The objective function is 

\begin{equation}
\max I(\beta, r ; \theta)-\lambda I(\beta, X ; \theta) \label{eqib}
\end{equation}
For the sake of simplicity, we omit the representation of subscripts. Due to the incomputability of $p (r | \beta)$ and $p (\beta)$, $q (r | \beta)$ and $h (\beta)$ are introduced for variational approximation. Thus, the final optimization objective is obtained as the variational lower bound of the original optimization objective. The lower bound of the first term of the objective function in Eq. (\ref{eqib}) is
\begin{equation}
\begin{aligned}
I(\beta, r) & \geq \int d r d \beta p(r, \beta) \log \frac{q(r \mid \beta)}{p(r)} \\
& =\int d r d \beta p(r, \beta) \log q(r \mid \beta)-\int d r p(r) \log p(r) \\
& =\int d r d \beta p(r, \beta) \log q(r \mid \beta)+H(r) \label{eqib2}
\end{aligned}
\end{equation}
The entropy $H(r)$ is irrelevant to the optimization objective, so it can be ignored. Furthermore, based on the Morkov chain and the edge probability density, Eq. (\ref{eqib2}) can be concluded by
\begin{equation}
I(\beta, r) \geq \int d z d r d \beta p(z) p(r \mid z) p(\beta \mid z) \log q(r \mid \beta)  \label{eqib3}
\end{equation}
For the variational approximation of the first term, we only need  samples from the joint data distribution and samples from the random encoder of the processable variational approximation $q (r | \beta)$.

For the second term, we use the same method to use $H(r)$ for variational approximation of $p (\beta)$, and obtain 
\begin{equation}
I(\beta, z) \leq \int d z d r d \beta p(z) p(r \mid z) p(\beta \mid z) \log \frac{p(\beta \mid z)}{h(\beta)} \label{eqib4}
\end{equation}
By combining the two terms in Eq. (\ref{eqib3}) and Eq. (\ref{eqib4}), a new variational lower bound can be obtained as 
\begin{equation}
\begin{gathered}
I(\beta, r)-\beta I(\beta, z) \geq \int d z d r d \beta p(x) p(r \mid z) p(\beta \mid z) \log q(r \mid \beta) \\
\quad-\lambda \int d z d r d \beta p(z) p(r \mid z) p(\beta \mid z) \log \frac{p(\beta \mid z)}{h(\beta)} \label{eqib5}
\end{gathered}
\end{equation}
The empirical estimate on the right-hand side of Eq. (\ref{eqib5}) is
\begin{equation}
\frac{1}{t} \sum_{t=1}^T\left[\int d \beta_{t-1} p\left(\beta_{t-1} \mid z_{t-1}\right) \log q\left(r_t \mid \beta_{t-1}\right)-\lambda p\left(\beta_{t-1} \mid z_{t-1}\right) \log \frac{p\left(\beta_{t-1} \mid z_{t-1}\right)}{h(\beta_{t-1})}\right]
\end{equation}
Combing Eq. (\ref{eq4}), the final objective function is 
\begin{equation}
\begin{gathered}
\min _{\theta, b, W} \sum_{t=1}^T\left\|r_t-\beta_{t-1} f_t\right\|^2 + \mathbb{E}\left[-\log q\left(r_t \mid g^\theta\left(z_{t-1}\right)\right)\right]\\+\lambda \operatorname{KL}\left[p\left(\beta_{t-1} \mid z_{t-1}\right), h(\beta_{t-1})\right]
\end{gathered}
\end{equation}
The prior distribution $h(\beta_{t-1})$ can be any distribution independent of $z_{t-1}$, so we simply chose the standard Gaussian distribution. 

\subsection{Optimization}
The pronounced non-linearity and non-convexity of neural networks, in conjunction with their extensive parameterization, render brute force optimization endeavors computationally demanding, often to the brink of infeasibility. A prevalent strategy to mitigate this challenge is the employment of stochastic gradient descent (SGD) \cite{amari1993backpropagation} for the training of neural networks. This approach diverges from conventional gradient descent, which computes gradients based on the entire training dataset at each iteration of the optimization process. In contrast, SGD approximates the gradients by sampling a small, randomly selected subset of the data at each iteration, thereby trading off the precision of the optimization for computational tractability. A pivotal hyperparameter within SGD is the learning rate, which dictates the magnitude of the descent steps. As the gradient approaches zero, it becomes imperative to correspondingly diminish the learning rate to zero; failure to do so may result in the noise inherent in the gradient computation overshadowing its directional cue. This treatise employs the adaptive moment estimation algorithm, an efficacious variant of SGD proposed by the work of \cite{kingma2014adam}. Adam algorithmically leverages the estimation of the gradient's first and second moments to dynamically adjust the learning rate for each parameter, thereby enhancing the convergence properties and robustness of the optimization process.

We employ ``batch normalization''\cite{santurkar2018does}, a straightforward technique designed to regulate the variability of predictive variables across different regions of the network and disparate datasets. The motivation behind this technique stems from the phenomenon of internal covariate shift, wherein the inputs to the hidden layers adhere to a distribution that differs from that of the corresponding layers in the validation samples. This issue is frequently encountered when fitting deep neural networks that involve a multitude of parameters and rather complex architectures. For each hidden layer (``batch'') at every training step, the algorithm performs cross-sectional demeaning and variance standardization on the batch inputs, thereby restoring the representational capacity of the units.

\section{Empirical Study}
\subsection{Data}
We conduct the experiments over US equity, the data is collected following the work of Autoencoder asset pricing models \cite{gu2021autoencoder}. We have constructed an extensive repository of stock-level predictive attributes, drawing upon the extant literature on the cross-sectional dynamics of stock returns. This compilation encompasses a total of $94$ distinct characteristics. Please see the work of \cite{gu2021autoencoder} or our supplementary materials for detailed data processing strategies.

We partition our dataset into three distinct temporal segments, ensuring the preservation of the sequential integrity of the data. The initial segment, referred to as the training subset (from 03/1957 to 12/1974), is utilized to estimate the model parameters under a specific configuration of tuning hyperparameters. These hyperparameters are pivotal to the efficacy of machine learning algorithms, as they govern the complexity of the model. The second segment, known as the validation subset (from 01/1975 to 12/1986), serves the purpose of hyperparameter optimization. We generate fitted values for the data points within the validation subset based on the model parameters estimated from the training subset. Subsequently, we compute the objective function, which is predicated on the errors derived from the validation subset, and engage in an iterative search for hyperparameters that maximize the validation objective. The selection of tuning parameters is informed by the validation subset, yet the parameters themselves are solely estimated from the training data. The validation process is designed to emulate an out-of-sample test of the model. Hyperparameter tuning essentially involves identifying a level of model complexity that is likely to yield robust out-of-sample performance. The fits from the validation subset are not strictly out-of-sample, as they are employed for tuning, which subsequently influences the estimation process. Consequently, the third segment, the testing subset (from 01/1987 to 01/2021), which is not utilized for either estimation or tuning, represents a true out-of-sample dataset. It is employed to assess the out-of-sample performance of the methodology.

\subsection{Evaluation Metric}
We evaluate our asset pricing model by using the out-of-sample $R^2$ and tangency portfolio Sharpe ratio. 
The $R^2$ quantifies the explanatory power of the concurrent realizations of factors, thereby evaluating the model's descriptive capacity regarding the risk profile of individual equities, calculated by
\begin{equation}
R_{total}^2=1-\frac{\sum_{t \in \mathcal{T}_{OOS}}\left(r_t-\beta_{t-1} f_t\right)^2}{\sum_{t \in \mathcal{T}_{OOS}} r_t^2}
\end{equation}

\begin{equation}
R_{pred}^2=1-\frac{\sum_{t \in \mathcal{T}_{OOS}}\left(r_t-\beta_{t-1} \hat{f}_{t-1}\right)^2}{\sum_{t \in \mathcal{T}_{OOS}} r_t^2}
\end{equation}
where $\mathcal{T}_{OOS}$  indicates the set including testing samples.  

We use out-of-sample tangency portfolio Sharpe ratio among factor portfolios to evaluate  the multi-factor mean-variance efficiency. The return of the tangency portfolio for a given set of factors is ascertained on an exclusively out-of-sample foundation, utilizing the mean and covariance matrix of the estimated factors from period $t$, and subsequently monitoring the return in the subsequent period $t+1$.

\subsection{Quantitative Results}
Table \ref{tab:my_label} reports the out-of-sample  $R^2$ results of each stock. We set the factor number $K$ to $1$, $3$, $6$, $12$, $24$, respectively. The compared method is Fama-French (``FF''), ``PCA'', ``IPCA'' and Autoencoder with one hidden beta layer of 32 neurons (``CA1''). We perform information bottleneck on both ``IPCA'' and ``CA1''. ``IPCA+IB'' signifies the incorporation of information bottleneck (``IB'') constraints into the objective function of ``IPCA'', which is then solved using the gradient descent method instead of using analytical methods. It is observed that when the number of factors is relatively small ($K=1,3,6$), the information bottleneck  methods do not enhance $R^2$ performance, and instead, it diminishes the out-of-sample $R^2$. This is attributed to the fact that with a limited number of factors, introducing irrelevant information constraints can impair model fitting. Conversely, as the number of factors increases($K=12,24$), methods without information bottleneck constraints exhibit deteriorating  $R^2$ performance, indicative of increasing overfitting. Our ``CA1+IB'' method, however, achieves superior results with an $R^2$  of 13.9, outperforming all other methods.

\begin{table}[]
    \centering
\begin{tabular}{l c c c c c}
\hline Model & K=1 & K=3 & K=6 & K=12 &  K=24\\
\hline FF & 4.5 & 3.0 & -6.2 & -10.2 & -11.4 \\
 PCA & 7.1 & 4.6 & 3.4 & 2.2 & 1.6 \\
 IPCA & 10.9 & 12.0 & 13.7 & 8.6 & 8.2 \\
 CA1 & 10.3 & 11.6 & 13.4 & 8.9 & 7.6 \\
\hline IPCA + IB & 8.5 & 11.2 & 13.3 & 10.5 & 9.9 \\
 CA1 + IB & 9.7 & 11.0 & 13.0 & 13.9 & 13.5 \\
\hline
\end{tabular}
    \caption{Out-of-Sample $R^2$(\%) Comparison}
    \label{tab:my_label}
\end{table}

We report out-of-sample factor tangency portfolio Sharpe ratios in Table \ref{tab:my_label2}. We can observe similar results to the $R^2$ evaluation from Table \ref{tab:my_label2}. The highest Sharpe ratio is obtained when using the Autoencoder with one hidden beta layer of 32 neurons under our information bottleneck constraint (``CA1+IB''). We conclude that as the number of factors grows, the information bottleneck methods yield more favorable outcomes without overfitting. 
\begin{table}[]
    \centering
\begin{tabular}{l c c c c c}
\hline Model & K=1 & K=3 & K=6 & K=12 &  K=24\\
\hline FF & 0.36 & 0.37 & 0.67 & 0.66 & 0.60 \\
 PCA & 0.21 & 0.17 & 0.24 & 0.36 & 0.33 \\
IPCA & 0.23 & 1.52 & 3.01 & 3.33 & 3.24 \\
 CA1 & 0.29 & 1.28 & 3.87 & 3.21 & 3.42 \\
\hline IPCA+IB & 0.12 & 1.04 & 2.98 & 3.62 & 3.63 \\
 CA1+IB & 0.20 & 1.35 & 3.01 & 3.90 & 3.94 \\
\hline
\end{tabular}
   \caption{Out-of-Sample Factor Tangency Portfolio Sharpe Ratios}
    \label{tab:my_label2}
\end{table}

\subsection{Qualitative Results}
\subsubsection{Visualization of Mutual Information }
\begin{figure}[htbp]
	\centering
	\begin{subfigure}{0.325\linewidth}
		\centering
		\includegraphics[width=0.9\linewidth]{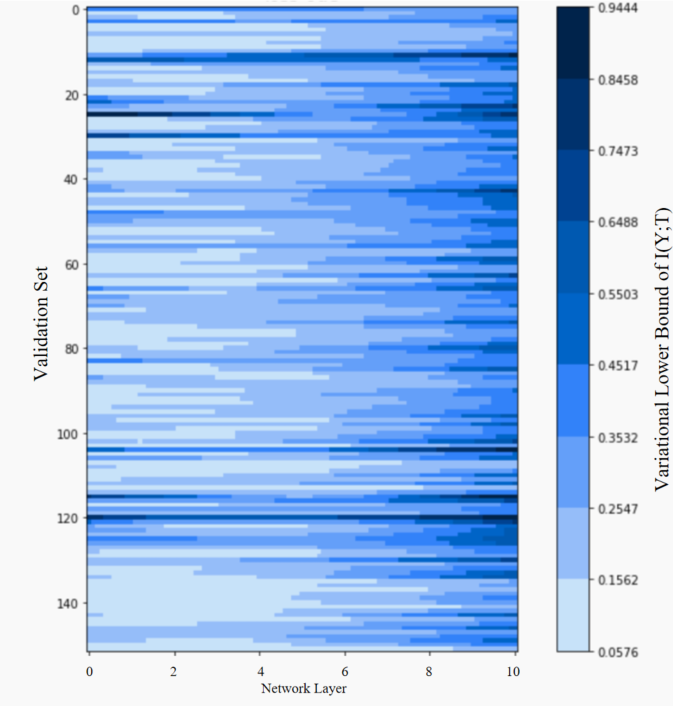}
		\caption{Validation Subset}
		\label{chutian3}
	\end{subfigure}
	\centering
	\begin{subfigure}{0.35\linewidth}
		\centering
		\includegraphics[width=0.9\linewidth]{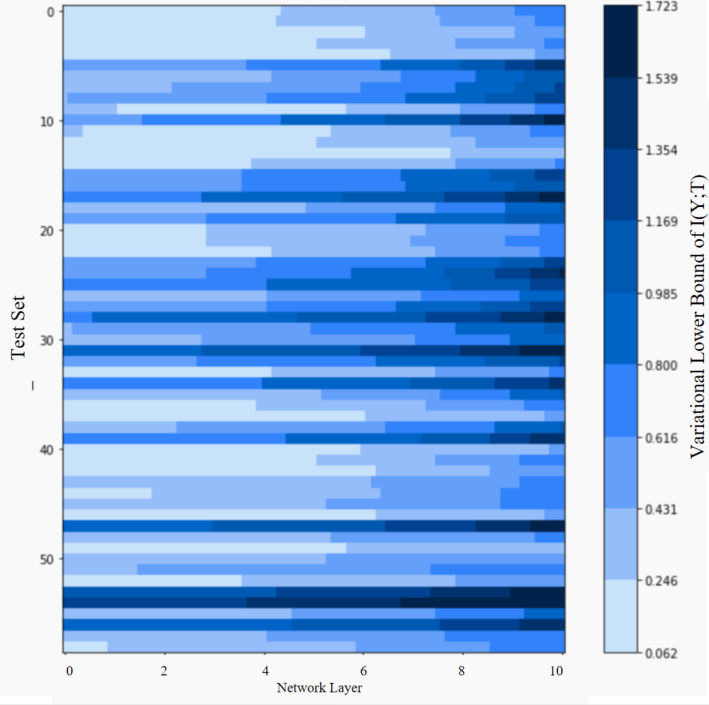}
		\caption{Testing Subset}
		\label{chutian3}
	\end{subfigure}
    \caption{Mutual information between the factor loadings $\beta$ from different network layers and the returns $r$}
	\label{fig1}
\end{figure}

In Figure \ref{fig1}, we visualized the mutual information in the validation subset and the last 5 years of testing subset. The horizontal axis represents the number of hidden layers of the factor loading network. Originally, ``CA1'' has only  one hidden layer, but now we  expand it to 10 layers. The vertical axis represents the mutual information $I(\beta,r)$ of different monthly frequency samples. On both the validation and test sets, it can be seen that as the number of network layers increases (i.e., the non-linear modeling ability improves), the mutual information gradually increases. That means that the last layer with the highest non-linear often has the largest mutual information, which verifies the effectiveness of mutual information constraints. It can also be seen that for some samples from certain months, a simple network can obtain significant mutual information, and these samples can be considered to have a relatively low signal-to-noise ratio. 

\subsubsection{Visualization of Cumulative Return}
\begin{figure}[htbp]
	\centering
	\begin{subfigure}{1\linewidth}
		\centering
		\includegraphics[width=1\linewidth]{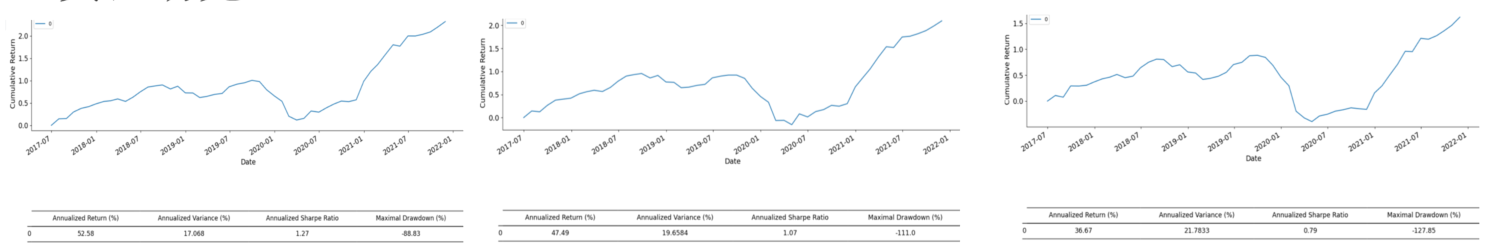}
		\caption{w/o Information Bottleneck}
		\label{chutian3}
	\end{subfigure}
    \qquad
	\centering
	\begin{subfigure}{1\linewidth}
		\centering
		\includegraphics[width=1\linewidth]{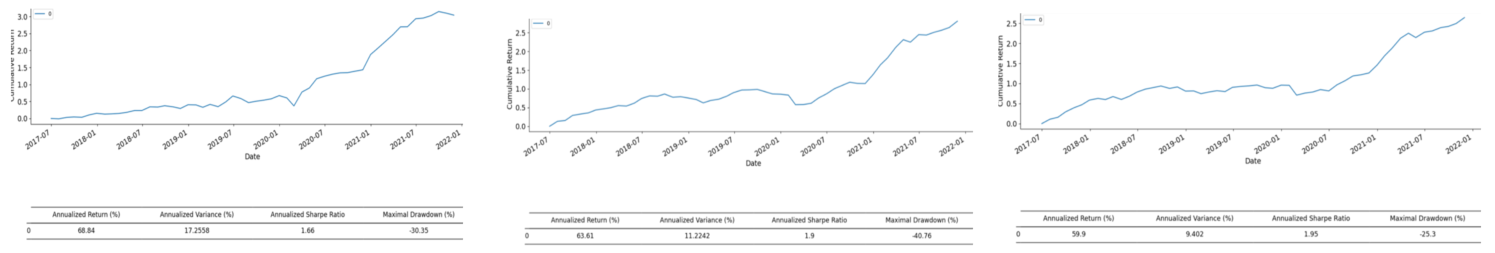}
		\caption{w/ Information Bottleneck}
		\label{chutian3}
	\end{subfigure}
    \caption{The curves of cumulative return in the last 5 years of testing subset. The number of hidden layers from left to right is 1, 3, and 6, respectively.}
	\label{fig2}
\end{figure}
Figure \ref{fig2}  shows the  curves of cumulative return in the last 5 years of testing subset. We set the factor loading network layer as 1, 3, and 6, and calculate the tangency weights to calculate out-of-sample returns. Without information bottleneck constraints (``w/o Information Bottleneck''), we find that the Sharpe ratio of the Autoencoder method is 1.27, 1.07, 0.79, which is decreased with the deepening of the network. Especially when the COVID-19 epidemic occurred in January 2020, the curve decreased significantly.  It is worth noting that Autoencoder itself uses LASSO-like sparse regularization to prevent overfitting, but it seems not work well. When using information bottleneck constraints, the Sharpe rate is also increasing with the deepening of the network. It can be considered that the model effectively fits the data rather than noise, When using a network with stronger nonlinear capability, and thus the  overfitting is less significant.

\section{Conclusion}
In this paper, we have explored the application of deep neural networks (DNNs) in financial asset pricing and identified the critical issue of overfitting to noise within financial data, which can significantly impair the performance of sophisticated models. To combat this challenge, we introduced an innovative Information Bottleneck Asset Pricing Model that is designed to navigate the complexities of financial data while mitigating the impact of noise. Our model employs a data compression strategy that selectively eliminates redundant information and preserves the essential components necessary for accurate asset pricing. By imposing mutual information constraints during the nonlinear mapping process, we ensure a progressive reduction in the mutual information between the input data and its compressed representation, while simultaneously increasing the mutual information between the compressed representation and the output prediction. This dual approach effectively filters out irrelevant noise without compromising the integrity of the final asset pricing. The Information Bottleneck framework allows our model to capitalize on the nonlinear modeling capabilities of DNNs to capture the intricate relationships within financial data. Moreover, it ensures that the model is not only adept at capturing these relationships but also robust against the adverse effects of noise during the information compression process. This results in a model that is both powerful in its predictive accuracy and resilient to overfitting. In conclusion, our Information Bottleneck Asset Pricing Model offers a significant advancement in the field of financial asset pricing. It provides a robust framework that leverages the strengths of DNNs while addressing their vulnerabilities to noise. By doing so, it paves the way for more reliable and effective asset pricing models that can withstand the challenges posed by complex and noisy financial data. This research not only contributes to the theoretical understanding of DNNs in asset pricing but also has practical implications for the development of more accurate and stable financial models.

\bibliographystyle{alpha}
\bibliography{sample}

\end{document}